\documentclass[prl,letterpaper,twocolumn,showpacs,superscriptaddress,floatfix]{revtex4}
\usepackage{graphicx,psfrag,amsmath,amssymb,amsfonts,bm,bbm,latexsym,color,dcolumn}

\begin{document}

\title{Thermalization in open classical systems with finite heat baths}

\author{S. Taylor Smith}

\affiliation{Department of Physics and Astronomy,Dartmouth
  College,6127 Wilder Laboratory,Hanover,NH 03755,USA}

\author{Roberto Onofrio} 

\affiliation{Dipartimento di Fisica ``Galileo Galilei",Universit\`a di
Padova,Via Marzolo 8,Padova 35131,Italy} 

\affiliation{Center for Statistical Mechanics and Complexity,INFM-CNR,Unit\`a di
  Roma 1,Roma 00185,Italy}

\affiliation{Department of Physics and Astronomy,Dartmouth
  College,6127 Wilder Laboratory,Hanover,NH 03755,USA}

\begin{abstract}
We discuss thermalization of a test particle schematized as 
a harmonic oscillator and coupled to a Boltzmann heat bath of finite size 
and with a finite bandwidth for the frequencies of its particles. 
We find that complete thermalization only occurs when the test particle frequency 
is within a certain range of the bath particle frequencies, and for a certain range of 
mass ratios between the test particle and the bath particles. These results have implications 
for the study of classical and quantum behaviour of high-frequency
nanomechanical resonators.
\end{abstract}


\maketitle

Since the pioneering studies on Brownian motion and its interpretation at the molecular level 
\cite{Einstein}, the study of open systems has been a crucial area in
both classical and quantum statistical mechanics. 
The theory of open systems allows one to bridge the closed, pure-state dynamics of a particle 
with the open, mixed-state dynamics in the presence of coupling to an environment. 
The deterministic dynamics of the particle considered as a closed system is replaced in the quantum 
regime by a stochastic Schr\"odinger equation, corresponding to a stochastic Newton 
equation in the classical limit. 
In the classical case, the environment can be represented as a heat
reservoir consisting of a large set of harmonic oscillators with
Boltzmann energy distribution, all symmetrically coupled to a specific
particle, which hereafter we will refer to as the {\sl test particle}. 
In the limit of an infinite number of oscillators, and by properly
choosing the density of states and the initial conditions, one can  
derive a stochastic equation which leads to solutions with 
Boltzmann energy distribution for the test particle, 
{\it i.e.} thermalization at the reservoir temperature.  

To the best of our knowledge, however, there have been no investigations of 
the conditions under which thermalization of a test particle occurs in the 
presence of a realistic heat bath with limited resources; namely a finite 
number of degrees of freedom and a finite bandwidth. The closest discussion 
in the literature regards thermalization of a test particle with an independent 
oscillator heat bath, but with oscillators distributed in a generic and unbound 
frequency range \cite{Ford}, the study of interacting Fermi systems with finite
numbers of particles \cite{Flambaum}, and the celebrated
Fermi-Pasta-Ulam problem, where nonlinear interactions are crucial \cite{FPU}. 
To consider a finite bandwidth for a finite number 
of oscillators making a heat bath is significant because in many contexts, in 
particular for systems of interest in mesoscopic physics and nanotechnology, the 
size of the environment is small and may no longer justify the thermodynamic limit. 
This leads naturally to an infrared cut-off for the density of states of the bath, 
as the latter cannot support wavelengths much larger than its spatial extent. 
Furthermore, the finite amount of energy in any realistic environment demands an 
ultraviolet cut-off in its density of states. In this paper, we describe results 
on the thermalization of a test particle in the presence of a heat bath comprising 
a finite number of harmonic oscillators with frequencies distributed within a 
finite bandwidth. The main result of our analysis is the emergence of
diverse scenarios for the statistical distribution of the energy of
the test particle, such as thermalization at the heat bath
temperature, Boltzmann distribution at a smaller temperature than the 
heat bath, or no thermalization, depending on the interplay between 
the intrinsic frequency of the test particle and those of the heat bath.  
While our study focuses on the classical case, it points to a broad
range of implications in the physics of nanomechanical structures, whose
displacements may be monitored at or beyond the standard quantum limit 
\cite{Braginsky,Caves}. Furthermore, some of the results reported here may find 
application wherever an open system approach is necessary,
encompassing systems as diverse as classical plasmas and
self-gravitating objects, ultracold trapped atoms and quasiparticles in nanostructures.

The starting point for our analysis is the Hamiltonian of the test
particle plus environment \cite{Ford1,Caldeira,Ford2}:
\begin{equation}
H_\mathrm{tot} = \frac{P^2}{2 M}+ \frac{1}{2} M \Omega^2 Q^2 + 
\sum_{n=1}^N \left[\frac{p_n^2}{2 m}+\frac{m \omega_n^2}{2} (q_n-Q)^2 \right].
\end{equation}
Here the test particle, with generalized coordinates $(Q,P)$, is modeled as a harmonic 
oscillator of mass $M$ and angular frequency $\Omega$. The $\mathrm{n}$-$\mathrm{th}$ particle 
in the reservoir, a harmonic oscillator of mass $m$ and angular
frequency $\omega_n$, is characterized by generalized coordinates $(q_n,p_n)$. 
To avoid unnecessary complications 
at this stage, the systems are assumed to be one-dimensional. 
The coupling between the test particle and the particles in the reservoir is chosen to be 
translationally invariant, which eliminates the appearance of renormalization terms \cite{Hakim,Patriarca}. 
By writing the Hamilton equations and solving for the particles of the heat bath, a generalized 
Langevin equation for the test particle is obtained \cite{Mori,Kubo}: 

\begin{equation}
\ddot{Q}(t) + \int_{t_0}^{t} ds \Gamma(t-s) \dot{Q}(s)+ \Omega^2 Q(t)=\Pi(t),
\end{equation}
 
\noindent
where the kernel of the dissipative term (non local in time) and the fluctuation 
force term are respectively:
\begin{eqnarray}
\Gamma(t-s) & = & \frac{m}{M} \sum_{n=1}^N \omega_n^2 \cos[\omega_n (t-s)], \\
\Pi(t) & = & \frac{m}{M} \sum_{n=1}^N \omega_n^2 \{
(q_n(t_0)-Q(t_0))\cos[\omega_n(t-t_0)] \nonumber \\ 
 && +\frac{p_n(t_0)}{m \omega_n} \sin[\omega_n(t-t_0)] \}. 
\end{eqnarray}

In the limit of infinite $N$  extending over the continuum frequency range 
$\omega_n \in [0, +\infty)$, with density of states $d N/d \omega \propto 1/\omega^2$, and if the initial 
conditions for the bath particles $(q_n(t_0), p_n(t_0))$ are chosen to realize a Boltzmann energy distribution, 
the test particle is described by a Langevin equation (for an extensive discussion see \cite{Presilla1})
\begin{eqnarray}
dQ(t) & = & P(t) dt / M, \\ 
dP(t) & = & -[\gamma P(t)+M \Omega^2 Q(t)] dt +  \delta dw(t),
\label{Langevin}
\end{eqnarray}
\noindent
where $\gamma= (\pi m \omega^2/2M)  (dN/d \omega)$ is the dissipation
factor (which becomes independent of $\omega$ for a density of
states $\propto 1/\omega^2$) and $\delta=\sqrt{2 m \gamma k_B T}$. 
The stationary solution of Eqs. (5)-(6) corresponds to the thermalization of the test particle energy 
with the same Boltzmann distribution as the bath. Such a situation is highly idealized, since it corresponds 
to an infinite number of oscillators with all possible frequencies, and these conditions are not always 
met in real physical systems.

To study the conditions under which thermalization occurs in a more realistic setting, we have 
investigated the dynamics of a reservoir characterized by a finite frequency spectrum 
$\omega_n \in [\omega_\mathrm{IR}, \omega_\mathrm{UV}]$, a finite number of oscillators $N$, a 
finite mass ratio $m/M$ between the masses of the bath particles and the test particle, for a uniform 
density of states, obtained by a random choice of the bath particle's frequencies, such that 
$d N/d \omega \propto \theta(\omega-\omega_\mathrm{IR})+\theta(\omega_\mathrm{UV}-\omega)-1$.

We do this by generating explicit numerical values for the parameters which describe each bath 
oscillator, solving the equations of motion from Eq. (1) for the motion of the test particle, 
and sampling the test particle energy over time. The initial conditions are chosen, for the 
bath oscillators, to be randomly distributed in phase space in such a way as to generate a 
Boltzmann distribution for their energy, and consequently a well defined temperature, while 
the test particle starts with zero total energy. The system of $N+1$ oscillators can then be 
solved exactly after numerical diagonalization of the matrix describing the Hamiltonian evolution.
Since the test particle is a single system, no meaningful definition of its temperature 
may be given in terms of the energy distribution of a statistical
ensemble at a given time. 
An effective temperature can still be defined by sampling the energy of the test
particle during its time evolution, collecting an energy
distribution over time, and fitting the energy distribution with a
Boltzmann distribution law. 
This operative definition does, however, preclude observation of
transient behaviour, at least on timescales smaller or comparable
to the time interval over which the energy distribution is collected, in the 
test particle temperature. Indeed, we find that the first effective
temperature we are able to define, when enough sample 
points have been collected to limit the statistical error, is already 
the equilibrium temperature.
This precludes in principle the study, for instance, of the dynamics of nonequilibrium stationary
states, nonequilibrium phenomena such as aging in glassy systems,
or the approach to thermal equilibrium.
Also, to avoid the effect of possible recurrences and spurious periodicities we have randomly 
chosen the sampling times for the measurement of the test particle energy.

The energy distribution of the test particle 
is shown in Fig. 1 for various values of the bandwidth $\Delta \omega= \omega_\mathrm{UV}-\omega_\mathrm{IR}$
of the frequency spectrum of the reservoir. 
For a test particle interacting with a finite number of oscillators all degenerate 
in frequency ($\omega_n =\omega_R$), we do not observe thermalization, as the resulting 
energy distribution is quite far from being Boltzmann in nature. Indeed, if the initial conditions 
for the bath oscillators are distributed symmetrically in phase space (such that 
$\sum_n q_n(t_0) \simeq 0$, $\sum_n p_n(t_0) \simeq 0$), Eqs. (3)-(4),
and (2) may be approximated 
respectively as:
\begin{eqnarray}
&\Gamma(t-s)= \xi \omega_R^2 \cos[\omega_R (t-s)], \\
&\Pi(t)= - \xi \omega_R^2 Q(t_0) \cos[\omega_R(t-t_0)], \\
&\ddot Q(t)+ \xi \omega_R^3 \int_{t_0}^t Q(s) \sin[\omega_R(t-s)] ds+\tilde{\Omega}^2 Q(t)=0,
\end{eqnarray}
where $\tilde{\Omega}=(1+\xi)^{1/2} \Omega$ and $\xi=Nm/M$.

\begin{figure}[t]
\includegraphics[width=1.00\columnwidth,clip]{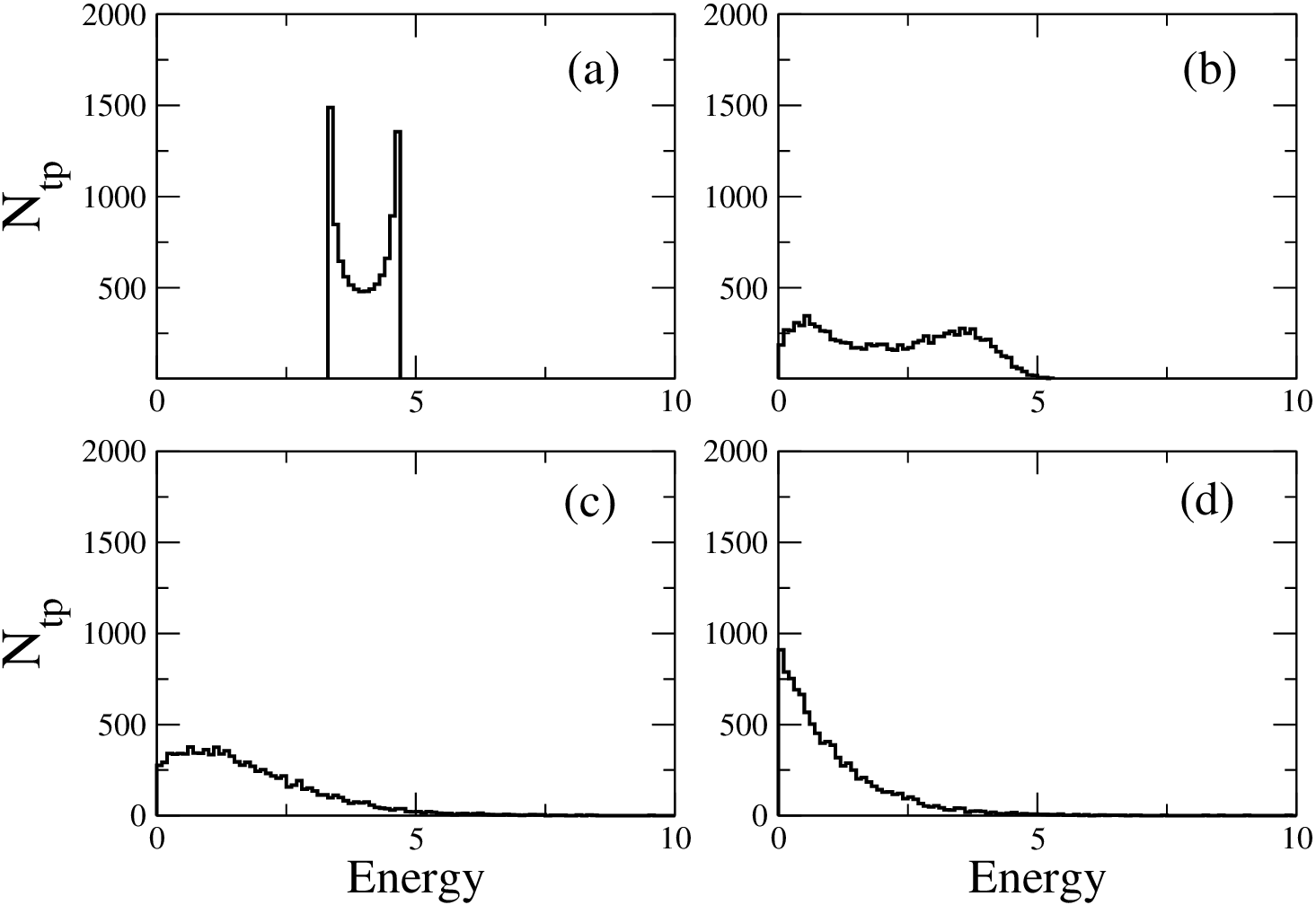}
\caption{Dependence of the energy distribution of the test particle on the frequency 
bandwidth of the bath oscillators. The number of times $N_{\mathrm{tp}}$ the
test particle is found to be with energy between $E$ and $E+\Delta E$
(with $\Delta E=0.5$ the energy binning interval) is plot versus the 
energy of the test particle (expressed in arbitrary units). The heat bath consists of 400 
harmonic oscillators with frequencies centered around the test particle frequency, 
$\Omega=60$ in arbitrary units, and with bandwidth equal to $\Delta \omega=$0 
(resonant bath, (a)), $\Delta \omega=$10 (b), $\Delta \omega=$40 (c),
and $\Delta \omega=$80 (d). 
The test particle energy, initially assigned to be zero, is sampled at random times, 
on average every 0.1 times its intrinsic period of oscillation. 
Apart from the degenerate bath case \cite{note1}, we have observed
that the test particle approaches its stationary energy state
regardless of its initial conditions provided that these correspond to an initial 
energy small enough. Measurements on the test particle are taken after
waiting a time long enough to yield stationary distributions, 
with a total of $10^4$ measurements. The mass ratio between the bath 
particles and the test particle is $m/M=10^{-3}$.} 
\label{fig1}
\end{figure}

\begin{figure}[t]
\includegraphics[width=1.00\columnwidth,clip]{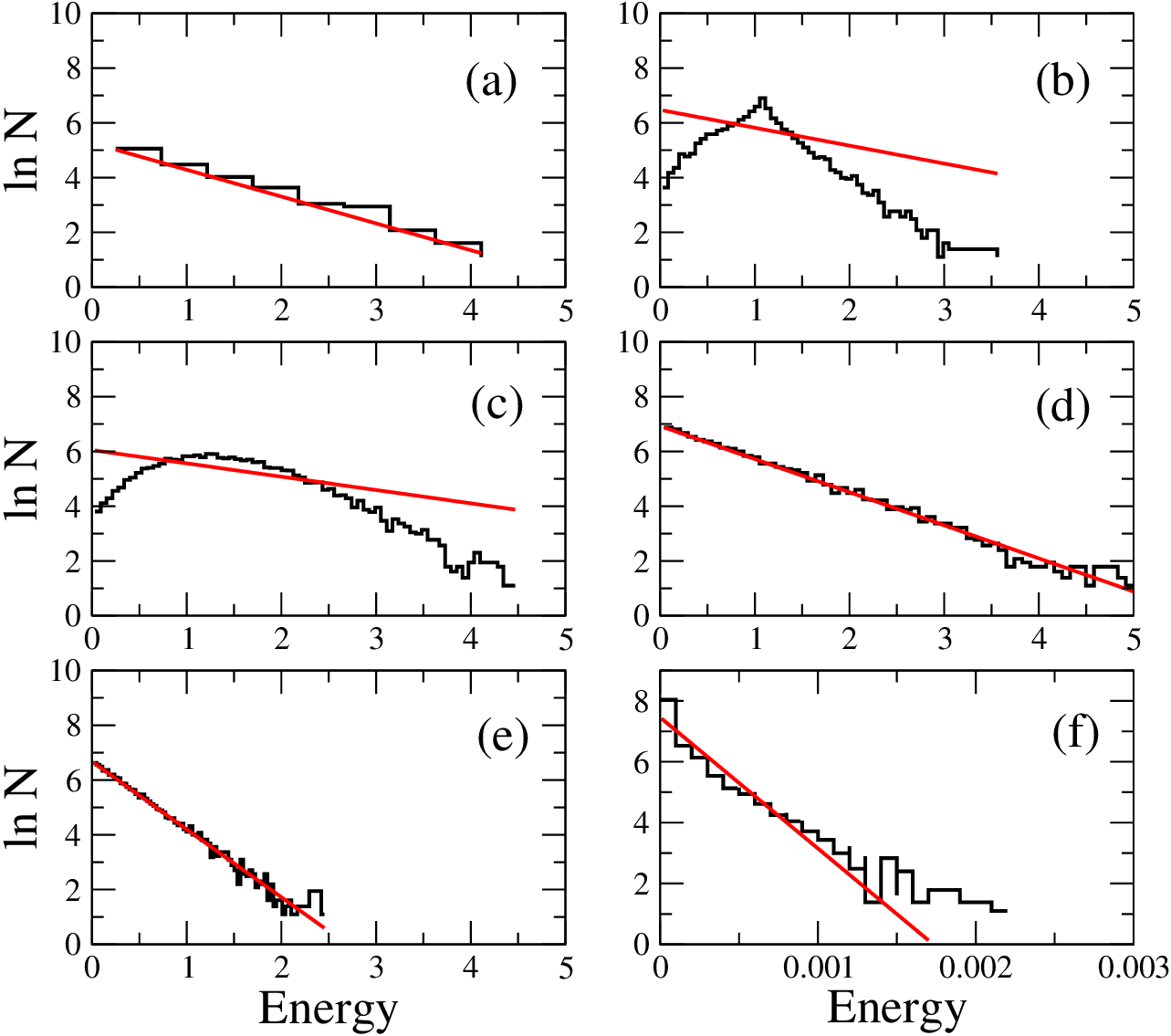}
\caption{Dynamics of thermalization. From top-left to bottom right: energy distribution 
for the heat bath composed of 400 oscillators (a), and energy
distribution of the test particle, after $10^4$ random samplings, with a frequency 
$\Omega=\omega_\mathrm{IR}/20$ (b), 
$\Omega=\omega_\mathrm{IR}$ (c), 
$\Omega=(\omega_\mathrm{IR}+\omega_\mathrm{UV})/2$ (d), 
$\Omega=\omega_\mathrm{UV}$ (e), and 
$\Omega=20\omega_\mathrm{UV}$ (f).   
The straight lines are best fits with Boltzmann distributions.
Since, as visible in (b) and (c), the energy distribution differs significantly 
from the Boltzmann distribution at small $\Omega$, the corresponding 
quoted temperatures from the Boltzmann fitting have to be interpreted
as effective temperatures, measuring the average energy of the test 
particle rather than the slope of a Boltzmann distribution.}
\label{fig2}
\end{figure}

The dynamics corresponding to Eq. (9) resembles that of multiple oscillators 
experiencing beating phenomena, with energy periodically transferred between the test 
particle and the heat bath with a harmonic law, as
$E_\mathrm{tp}(t)=E_0 \{ 1+\xi/2 [1- \cos (\tilde{\Omega} t/2)]\}$,
where $E_0$ is the bath's initial energy. 
The corresponding probability density $P(E)$ for the test particle to have 
energy between $E$ and $E+dE$ is $P(E) \propto (dE/dt)^{-1} \propto \{(E-E_0)[E_0(1+\xi)-E]\}^{-1/2}$.  
This behaviour is shown in Fig. 1a; evidently a monochromatic reservoir is too deterministic 
to allow scrambling of the exchanged energy, and we see here that the test 
particle does not thermalize. With even a modest bandwidth for the reservoir 
the coherent driving of the test particle does not take place at the same level, 
as shown in the top-right and bottom-left plots for increasing values of the 
bandwidth, and above a given value the energy distribution eventually approaches 
the Boltzmann distribution, as shown in Fig. 1d. 

These results are obtained by keeping the spectrum of the reservoir centered 
around the frequency of the test particle. An asymmetric situation, in which  
the oscillation frequency of the test particle is not centered 
within the heat bath bandwidth, is depicted in Fig. 2. 

One can obtain situations with no thermalization, thermalization at a temperature 
different from the one of the reservoir, or complete thermalization, depending 
upon the relative location of the test particle frequency with respect to the frequency spectrum 
of the reservoir. The reservoir, whose energy distribution is depicted in the upper-left panel, 
is chosen with a bandwidth equal to the one in Fig. 1d. In the other five plots in Fig. 2, the 
energy distribution of the test particle is shown for increasing values of its oscillation frequency. 
In particular, at $\Omega < \omega_\mathrm{IR}$ (panels (b) and (c)),
the test particle does not thermalize, and its energy distribution
shows a peak at a finite energy. 
In fact, in the limit $\Omega \rightarrow 0$ (free test particle) and for $\Delta \omega \rightarrow 0$ 
(resonant bath) the equations of motion correspond to a driving force and an 
unlimited  increase of energy in time. The presence of a finite $\Delta \omega$ introduces 
dephasing among the oscillators and consequently a finite and stationary energy peak proportional 
to $\xi$. In panel (d), the particle reaches a Boltzmann distribution
with a temperature close to the one of the reservoir $T_\mathrm{tp}
\simeq T_\mathrm{res}$, whereas in panels (e) and (f) the Boltzmann
distributions of the test particle correspond to significantly lower temperatures. 

\begin{figure}[t]
\includegraphics[width=1.00\columnwidth,clip]{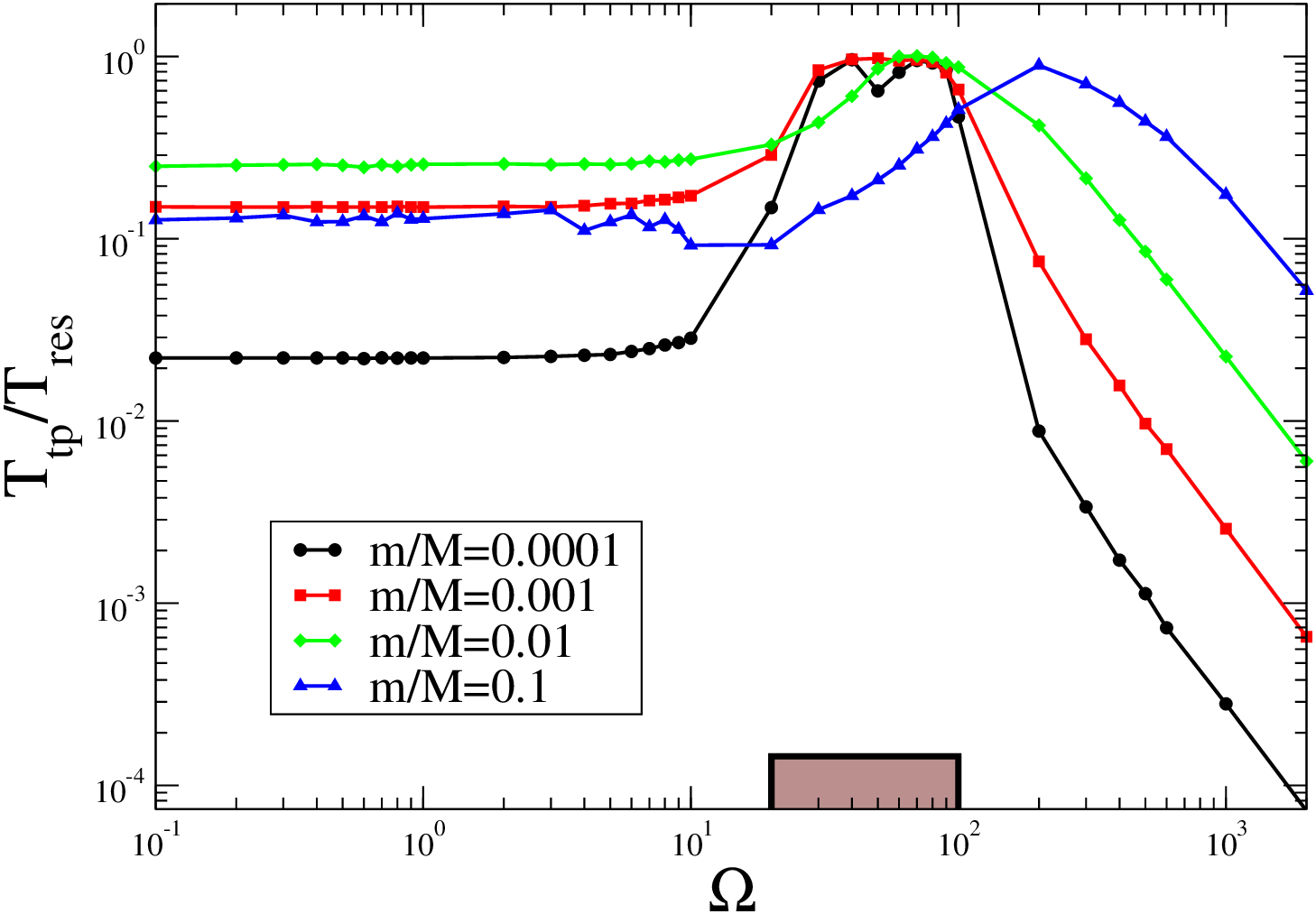}
\caption{Efficiency of thermalization versus test particle frequency.
Shown is the ratio between the temperature of the test particle and 
the temperature of the heat bath versus the angular frequency of 
the test particle (in arbitrary units), for different mass ratios
$m/M$. The heat bath is composed of 400 oscillators with uniform density of states 
between $\omega_\mathrm{IR}$=20 and $\omega_\mathrm{UV}$=100, as
indicated by the brown rectangle in the lower part of the plot.}
\label{fig3}
\end{figure}

By repeating the procedure for different values of the angular frequency 
of the test particle, we find three different regimes, as shown in Fig. \ref{fig3}. 
At low angular frequencies $\Omega << \omega_\mathrm{IR}$, the test particle approaches 
an equilibrium state of non-Boltzmann nature corresponding to an effective temperature 
significantly lower than the reservoir temperature, with a flat dependence upon the frequency. 
The lack of thermalization in this situation resembles the
  situation of the Brownian motion of a free particle 
(obtainable in our case as the limit $\Omega \rightarrow 0$) 
coupled to a bath with super-Ohmic density of states \cite{Grabert}.  
In the intermediate regime  $\omega_\mathrm{IR} \leq \Omega \leq  \omega_\mathrm{UV}$, the test particle 
generally approaches complete thermalization at $T_\mathrm{tp} \simeq
T_\mathrm{res}$. In the region where $\Omega >> \omega_\mathrm{UV}$,
the test particle has a Boltzmann energy distribution at a temperature 
$T_\mathrm{tp}<< T_\mathrm{res}$ which is frequency-dependent.  
This may be understood by considering the fact that a finite ultraviolet cut-off for 
the heat bath implies non-Markovian behaviour. Then a high-frequency test particle should 
be dynamically decoupled from the heat bath in analogy to the mechanism introduced 
in \cite{Viola,Vitali} for open quantum systems.
Notice that, in the case of $m/M=0.1$ in Fig. 3 (and, less manifestly, for 
the curve at $m/M=10^{-2}$), the region of nearly complete
thermalization is shifted towards significantly 
larger frequencies with respect to the interval $[\omega_\mathrm{IR},
\omega_\mathrm{UV}]$.  The obtained data are stable with respect to the
choice of the initial time at which the sampling of the test particle
energy is performed; even waiting more than four orders of magnitude
does not change the thermalization pattern.

This behaviour may be qualitatively understood by thinking of the case of a degenerate heat bath.
In this extreme situation, the particle is either in resonance with the bath oscillators, 
trading energy in an efficient way, or otherwise always detuned,  
limiting the energy exchange both in amplitude and speed. 
Since the effective frequency of oscillation of the test particle in Eq. (9) is 
renormalized as $\tilde{\Omega}= (1+\xi)^{1/2} \Omega$, if $\xi$ is 
significantly larger than unity we expect the resonance to occur at angular frequencies 
higher than the intrinsic angular frequency of the test particle.
In the cases of $m/M=10^{-2}$ and $m/M=10^{-1}$ we expect, respectively, 
$\tilde{\Omega}= \sqrt{5} \Omega$ and $\tilde{\Omega}=\sqrt{41} \Omega$, 
and the shifts of the thermalization regions are in agreement with this scaling. 
This has as a further implication a non monotonic dependence 
of the thermalization upon the mass ratio $m/M$, which cannot be trivially 
addressed in a perturbative approach \cite{Plyukhin}. 
At constant $\Omega$, we expect that by increasing the mass ratio, the 
parameter $\xi$ becomes so large as to detune the effective response of 
the test particle, making its thermalization to the heat bath less efficient.

\begin{figure}[t]
\includegraphics[width=1.00\columnwidth,clip]{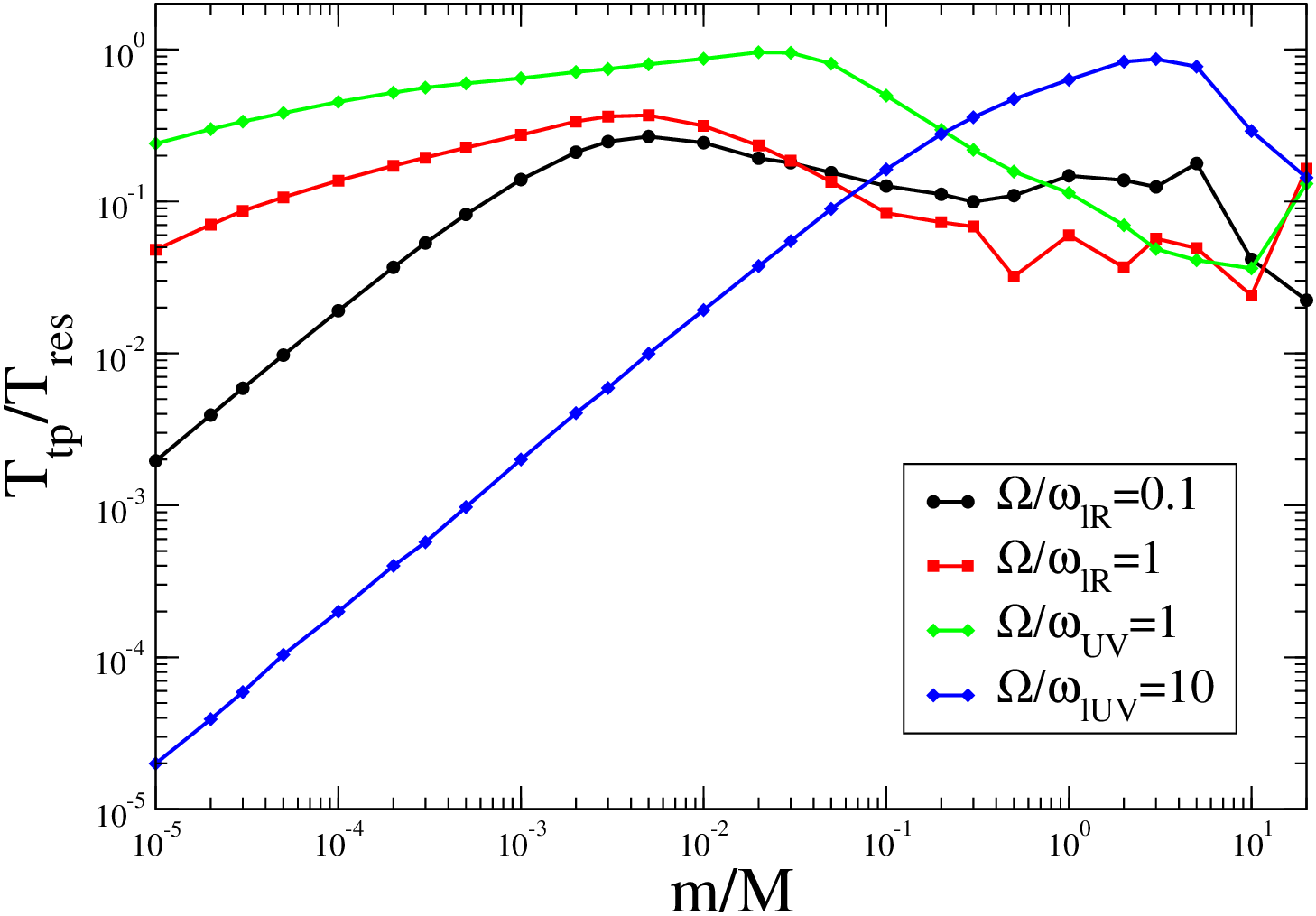}
\caption{Dependence of the efficiency of thermalization upon the particle masses.
Shown is the ratio between the temperature of the test particle and the temperature 
of the heat bath versus the ratios between the mass of the test particle and the mass 
of the oscillators in the heat bath. Different plots correspond to different 
frequencies of the test particle, from $\Omega=0.1 \omega_\mathrm{IR}$ to 
$\Omega=10 \omega_\mathrm{UV}$, as described in the legend.}
\label{fig4}
\end{figure}

This feature is clearly evidenced in Fig. \ref{fig4}, where the mass 
ratio dependence is shown for four frequencies in different spectral regions. 
For small values of $\Omega$, thermalization is initially proportional 
to $m/M$, then decreases when this parameter is in the $10^{-2}$ range. 
For $\Omega >> \omega_\mathrm{UV}$, the behaviour of $T_\mathrm{tp}/T_\mathrm{res}$ 
is well described by a direct proportionality law, {\it i.e.} 
$T_\mathrm{tp}/T_\mathrm{res} \simeq m/M$. 
We have also repeated the simulations for various examples of the density
of states of the particles in the heat bath, with very little
effect on the thermalization features apart from the peak region, as 
will be reported elsewhere \cite{note2}.

In conclusion, we have studied thermalization of a test particle in contact with 
a finite reservoir, showing that this occurs only for specific relationships 
between the frequency and the mass of the test particle and the corresponding 
quantities of the particles constituting the reservoir. These results may be 
understood as the intermediate case between the two extreme situations of 
a reservoir with an infinite bandwidth and a reservoir with a degenerate spectrum 
of frequencies. The analysis can be extended to generic, nonlinear couplings 
between the test particle and the reservoir with numerical techniques. 

The interpretation of two recent experiments in nanomechanics may be 
reconsidered in light of the results obtained here. 
In particular, the observation of discontinuous features in the
mechanical transfer function of high-frequency nanoresonators operated
at low temperatures reported in \cite{Mohanty}, could be attributed to 
non-equilibrium features of the test particle, in this case a high-frequency mode of a
nanomechanical structure. Our conjecture is that for nanomechanical structures the 
infrared cut-off frequency $\omega_\mathrm{IR}$ (ultraviolet cut-off 
frequency) is significantly larger (smaller) than in the bulk. 
The infrared cut-off is determined by the lowest frequency which sustains stationary 
vibrations in the structure, and for typical sound speeds ($v_s \simeq 10^3-10^4$ m/s) 
and typical sizes involved in high frequency nanoresonators (100 nm-1$\mu$m) as in \cite{Mohanty}, 
we can easily have $\omega_\mathrm{IR} \simeq $ 1-10  GHz, realizing a situation where 
$\Omega \leq \omega_\mathrm{IR}$. Likewise, the Debye frequency could differ significantly
with respect to the bulk situation due to the effective lower dimensionality of the nanostructure. 
The effective temperature felt by the nanoresonator could then be quite different from the one 
expected by measuring the external bath to which it is coupled.
This should be confirmed by both dedicated experiments on thermal properties, such as specific 
heat and heat conductivity, {\it ab initio} calculations of the
density of states, as well as repetitions of the experiments with 
higher frequency resonators and modified size of the structures, possibly 
leading to a new route to investigate non-equilibrium quantum
statistical mechanics \cite{Presilla2}. 

Also, in \cite{Schwab}, a phenomenon of cooling has been interpreted as 
due to quantum back-action of the read-out system, even though the 
system was well above the standard quantum limit.  
A more economical interpretation is instead available by imagining the 
effective temperature of the nanoresonator as resulting from the competition 
between two effective classical heat baths at different temperatures. 
This is in line with the phenomenon of cold damping introduced eight decades 
ago \cite{Ornstein}, and demonstrated for macroscopic resonators in 
\cite{Hirakawa}.  Moreover, the data show intriguing anomalies, since the 
effective noise temperature and the noise spectral densities reported 
give  values $10^2$ and 15 times above the standard 
quantum limit, respectively \cite{Schwab}. This large discrepancy could be due 
to a partial thermalization of the nanoresonator in the presence of two 
competing baths.
From our perspective, the study of the thermalization 
of a particle in simultaneous interaction with two heat baths should 
shed light on this phenomenon already at the classical level, especially 
focusing on the energy distribution which, in the presence of finite 
resources, is not necessarily of Boltzmann nature. 

\vspace{0.5cm}

We are grateful to Paolo Ricci for useful suggestions on the computer
codes, and to Lorenza Viola for a critical reading of the manuscript.  
STS has been partially supported by the Mellam Foundation at Dartmouth, and RO 
acknowledges partial support from the Italian MIUR under PRIN 2004028108$_{-}$001.


\begin{thebibliography}{99}
\bibitem{note1} In the case of the degenerate bath resonant with the
test particle the test particle trajectory depends strongly on the set
of initial conditions necessary to describe its non-Markovian evolution.
\bibitem{note2} It is worthwhile to note that the thermalization curves in Fig. 3 are all shifted 
towards higher values of $\Omega$ with respect to the frequency distribution of the heat bath. 
This is due to the larger weight of the higher frequencies (since the coupling strength is 
proportional to $\omega^2$ in Eqs. (3)-(4)). We have verified that by using the density of 
states $\propto 1/\omega^2$, this feature is removed.
\end{thebibliography}
\end{document}